\documentclass[aps,pre,twocolumn,preprintnumbers,amsmath,amssymb,floatfix,superscriptaddress]{revtex4-1}
\usepackage{amsmath,amsbsy,amssymb}
\usepackage[dvips]{graphicx}
\usepackage{xcolor}

\DeclareGraphicsExtensions{.eps,.ps}

\usepackage{bm}% bold math

\def\vec#1{{\ensuremath{\bm{#1}}}}
\def\half{{\textstyle \frac{1}{2}}}

\def\nl{\hfil\break}

\def\d{\textrm{d} }

\def\s#1{_\textrm{#1} }

\def\e#1{\textrm{e}^{#1} } %Euler e
\def\ber{\begin{eqnarray}}
\def\eer{\end{eqnarray}}
\def\be{\begin{equation}}
\def\ee{\end{equation}}
\def\beno{\begin{equation*}}
\def\eeno{\end{equation*}}
\def\bea{\begin{eqnarray}}
\def\eea{\end{eqnarray}}

\begin{document}
%\draft \twocolumn[\hsize\textwidth\columnwidth\hsize\csname
%@twocolumnfalse\endcsname

\title{Deep optical penetration dynamics in photo-bending}

\author{Daniel Corbett}
\affiliation{Manchester Institute of Biotechnology, University of Manchester, 131 Princess Street, Manchester M1 7DN, United Kingdom}
\author{Chen Xuan}
\affiliation{Department of Mechanics and Engineering Science, Fudan University, Shanghai 200433, China}
\affiliation{Cavendish Laboratory, University of Cambridge, 19 JJ Thomson Avenue, Cambridge CB3
0HE, United Kingdom}
\author{Mark Warner}
\affiliation{Cavendish Laboratory, University of Cambridge, 19 JJ Thomson Avenue, Cambridge CB3
0HE, United Kingdom}
%\email{mw141@cam.ac.uk}
\date{\today}
%%%%%%%%%%%%%%%%%%%%%%%%%%%%%%%%%%%%%%%%%%%%%%%%%%%%%%%%%%%%%%%
\begin{abstract}
We model both the photo-stationary state and dynamics of an illuminated, photo-sensitive, glassy liquid crystalline sheet. To illustrate the interplay between local tilt $\theta$ of the sheet, effective incident intensity, curvature and dynamics, we adopt the simplest variation of local incident light intensity with angle, that is $\cos\theta$. The tilt in the stationary state never overshoots the vertical, but maximum curvature could be seen in the middle of the sheet for intense light. In dynamics, overshoot and self-eclipsing arise, revealing how important moving fronts of light penetration are.  Eclipsing is qualitatively as in the experiments of Ikeda and Yu (2003).

\end{abstract}
%%%%%%%%%%%%%%%%%%%%%%%%%%%%%%%%%%%%%%%%%%%%%%%%%%%%%%%%%%%%%%%
%\vspace{0.2cm}
% \narrowtext
\maketitle
 %\begin{multicols}{2}
%%%%%%%%%%%%%%%%%%%%%%%%%%%%%%%%%%%%%%%%%%%%%%%%%%%%%%%%%%
\section{Introduction}\label{sect:intro}
\begin{figure}[!b]
\includegraphics[width=0.4\columnwidth]{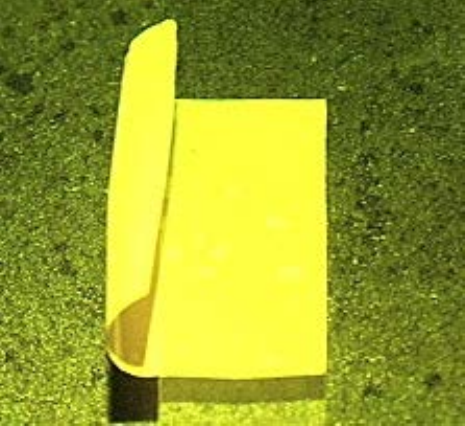}
\caption{A nematic sheet of Ikeda and Yu bending in response to illumination from above and self-eclipsing as the deformation develops. The right hand half is stuck to the support.}\label{fig:still}
\end{figure}

Nematic networks change their shape when their orientational order is induced to change thermally or, if dye molecules are present, optically. Mechanical strains can be between several hundred \% for elastomers and e.g. 4-5\% for glassy networks. We concentrate on the latter since they are strong (moduli $\sim 10^9-10^{10}$ Pa) and their directors immobile, eliminating further causes of mechanical response.
The experiments of Ikeda and Yu \cite{Ikeda:03} show two uniquely interesting new phenomena:\nl
(i) Large, optically-driven response in the direction of the polarization of light. Samples are polydomain so the only definers of directions are $\vec{k}_0$, the incident light's wave vector, and $\vec{E}_0$, the incident light's electric (and polarization) vector. At once, both the ease of delivery of stimulus and the control of mechanics by optics were demonstrated.
These aspects have been explored by many authors \cite{Finkphoto,Corbett:06,Mahadevan04,Corbett:08,Cviklinski:03,White:08}.\nl
(ii) Curling, in the direction of $\vec{E}_0$, occurred with large amplitude. Remarkably, curling continued so that much of the photo-glass sheet eclipsed itself; see Fig.~\ref{fig:still}. In shadow, one might expect curling to cease, or indeed reverse since the initially lower side of the sheet is now uppermost and being irradiated.
No explanation of this seemingly paradoxical phenomenon was advanced by \cite{Ikeda:03} or by subsequent authors. Our theory here shows that the phenomena uncovered by these seminal experiments reveal much about the non-linear and dynamical processes behind nematic photo-solid absorption of light and mechanical response. We suggest further experiments.

\section{Absorption, photomechanics and bend actuation}\label{sect:photomechanics}
Dye molecules are linear in their ground ({\it trans}, t) states and bent when excited ({\it cis}, c) by photon absorption. The number fraction of {\it cis}, $n\s{c}=1-n\s{t}$, increases by illumination, $I(x)$, and decreases by thermal recovery to $n\s{t}$ at a rate $1/\tau$, where $\tau$ is the c-lifetime, and where $I(x)$ is the intensity (Poynting flux) at depth $x$ into the sheet/cantilever. Thus
\be
\tau\frac{\partial n\s{c}}{\partial t}=\frac{I}{I_m}n\s{t}-n\s{c}\equiv\dot n\s{c}=-(\alpha I/I_0+1)n\s{c}+\alpha I/I_0,\label{eq:dynamics}
\ee
where $I_m$ is a material parameter $1/I_m =\Gamma\tau$ and $\alpha=I_0/I_m \equiv \Gamma I_0/(1/\tau)$ \cite{Corbett_PRL:07}. The constant $\Gamma$  subsumes an absorption cross section per chromophore and a quantum efficiency, while the reduced time $t/\tau$ derivative $\partial/\partial(t/\tau)$ is denoted by $\cdot$ .  We neglect c-absorption, background absorption and scattering in order to simply establish the qualitative aspects of the Ikeda and Yu phenomenon, which we succeed in doing.  Thus such additional sources of absorption are seemingly not central, and our model evidently has the essence of this mysterious effect.

The parameter $\alpha$ measures the ratio of the forward rate $\Gamma I_0$ (using the surface light intensity) to the back rate $1/\tau$. Large $\alpha$ implies strong perturbation from $n\s{c}=0$. Small $\alpha$ is the Beer limit where $n\s{c}\simeq0$ and absorption is by a dye population little perturbed from the dark state. We show that the Ikeda and Yu experiments reveal non-linearity ($\alpha\gtrsim1$) is vital. The photo-stationary state, $\dot n\s{c}=0$, gives
\be
n\s{c}=\frac{\alpha\mathcal{I}}{1+\alpha\mathcal{I}},\label{eq:stationary}
\ee
where $\mathcal{I}(x)=I(x)/I_0$ is an intensity at depth $x$ reduced by the intensity of light just having entered (at $x=0$).

Intensity is reduced with depth by the photon absorption in eq.~(\ref{eq:dynamics}) above, leading to $n\s{t}\rightarrow n\s{c}$
\be
\frac{\partial I}{\partial x}=-\frac{n\s{t}}{d}I,\label{eq:light absorption}
\ee
where the Beer Length $d$ subsumes cross sections, number densities {\it etc.}, and absorption depends on the number fraction $n\s{t}$ of absorbers. When $n\s{t} =1$, ($\alpha \ll 1$), the Beer limit $I(x)=I_0\e{-x/d}$ is obtained. We require finite conversion to get a mechanical response at all and to obtain the observed dynamics. Hence (\ref{eq:light absorption}) must be solved in the non-linear limit of $n\s{c}(x)\neq0$, that is, $n\s{t}$ is a function of $I$ itself \cite{Corbett_PRL:07}, either statically,  eq.~(\ref{eq:stationary}), or dynamically, eq.~(\ref{eq:dynamics}).

Creation of {\it cis} isomers lowers order and gives a photo-contraction along $\vec E$, that is a strain $\epsilon\s{p}=-Cn\s{c}$ in its simplest form, with $C$ a dimensionless scaling. For $\epsilon\s{p}\sim -0.04$ and $n\s{c}\sim 0.8$ (say), then $C\sim 1/20$. An $\epsilon\s{p}(x)$ varying with depth gives curvature $1/R$ as the solid aims to reduce the elastic cost of deviating from its new, natural local shape. The effective strain is
\be
\epsilon(x)=\frac{x}{R}+K-\epsilon\s{p}.
\ee
% hereafter absorbing $C$ into $1/R$ and redefining the constant $K$.

The longitudinal stress at a depth $x$ in the sheet is a modulus times this strain. Integrating the stress and the moment of the stress through the thickness, $w$, of the sheet to get the force and the torque, setting both these to zero, and cancelling the modulus yields the two equations \cite{Corbett_PRL:07}
\be
0=\int_{0}^{w}[\frac{x}{R}+K+ Cn\s{c}(x)]\d x=\int_{0}^{w}\frac{x}{R}+K+ C n\s{c}(x)]x\d x.
\ee
Eliminating between these two equations for $w/R$ one obtains
\be
\frac{w}{R}=\frac{12C}{w^2}\int_{0}^{w}\left(\frac{w}{2}-x\right)n\s{c}(x)\d x.\label{eq:curvature}
\ee
These equations must hold generally, even in a dynamically evolving system, in a limit where inertia can be ignored, for instance in the creeping motion seen by \cite{Ikeda:03}.

\section{Photo-stationary dye populations and mechanical response}\label{sect:stationary}

Using the stationary population (\ref{eq:stationary}) in the form $n\s{t}=1-n\s{c}=1/(1+\alpha\mathcal{I})$ in (\ref{eq:light absorption}) for $\mathcal{I}$ gives $\partial\mathcal{I}/\partial(x/d)=-\mathcal{I}/(1+\alpha\mathcal{I})$.
% where $'\equiv\frac{\partial}{\partial(x/d)}$.
Integration gives \cite{Statman:03,Corbett_PRL:07}
\be
\ln(\mathcal I(x))+\alpha({\mathcal I(x)}-1)=-x/d.\label{eq:productlog}
\ee
The solution for ${\mathcal I(x)}$ is in terms of the Lambert-W function (or ProductLog function), $W(c)$, which satisfies $c = W(c)\e{W(c)}$. Thus ${\mathcal I(x)}= \frac{1}{\alpha} W(\alpha \e{\alpha -x/d})$.  For large $\alpha$ -- intense light giving a high forward ${\it t}\rightarrow{\it c}$ rate compared with the decay rate $1/\tau$ -- the penetration is linear and very deep, $\mathcal I\sim1-x/(\alpha d)$ rather than exponential, ${\mathcal I}=\e{-x/d}$, which accounts for substantial mechanical response \cite{Corbett_PRL:07,Huo:10,Huo:11} when one would otherwise expect little response due to only a thin skin $d\ll w$ being Beer-penetrated. Dynamics will turn out to be strong evidence for non-linear effects.

Taking the solution (\ref{eq:productlog}) for $\mathcal I$ into $n\s{c}=\alpha\mathcal{I}/(1+\alpha\mathcal{I})$ and putting $n\s{c}(\mathcal I(x))$ into eq.~(\ref{eq:curvature}), then integrating using a variable change $\d (x/d) =-(1/\mathcal I+\alpha)\d\mathcal I$, and finally eliminating for $1/R$ gives \cite{Mahadevan04,Corbett_PRL:07}:
\be
\frac{(w/D)}{R}=\alpha\!\left[\frac{w}{d}\mathcal{I}_w-(1-\mathcal{I}_w)(1-\frac{w}{2d})-\frac{\alpha}{2}(1-\mathcal{I}_w^2)\right]\label{eq:curvature_Iw}
\ee
where the dimensionless combination $D=12C\!\!\left(\frac{d}{w}\right)^{\! 2}$ sets a scale to the reduced curvature $w/R$. (For $C=1/20$ and $w/d = 3$ of our illustration, $D=1/15$.)  Recall the definition of the reduced intensity, $\mathcal{I}$, after eq.~(\ref{eq:stationary}).
The solutions of (\ref{eq:productlog}) for $x = w$ are injected in this curvature expression.
The curvature $1/R$ increases as incident intensity, measured by $\alpha$, increases, but eventually must decrease again for very high $\alpha$ -- penetration is deep and $\mathcal{I}(w)\sim1-w/(\alpha d) \rightarrow 1$. The {\it cis} fraction $n\s{c}$ saturates to a high value (dye is depleted) and the consequent small variation of photo-strain $\epsilon\s{p}(x)$ with depth cannot induce bending.

\begin{figure}[h]
\includegraphics[width=0.8\columnwidth]{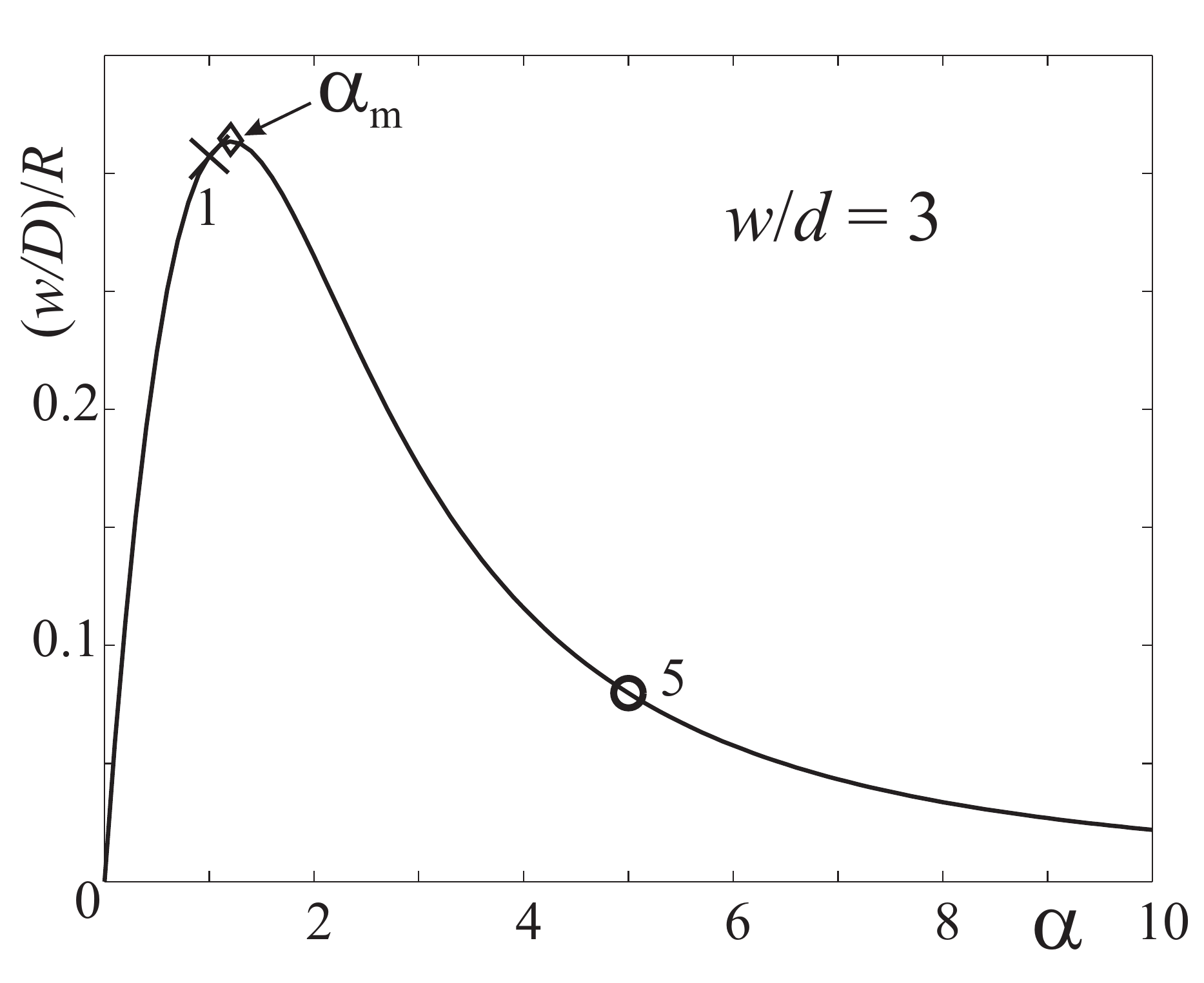}
\caption{Curvature as a function of incident light intensity $\alpha$ for $w/d=3$, with the optimal intensity, $\alpha\s{m}$ at $\diamondsuit$, for this thickness indicated. Two intensities $\alpha = 1, 5$ ($\times,\circ$) are used illustratively below.}\label{fig:curvature-light}
\end{figure}

Fig.~\ref{fig:curvature-light} shows the non monotonic function of curvature against incident light intensity $\alpha$: When incident light is very weak, the {\it cis} fraction is small thereby producing a small curvature. When incident light is very strong and therefore deeply penetrating, the {\it cis} fraction is close to 1 and nearly uniform through the thickness, and thus again there is hardly any curvature. Curvature is maximized at an optimal intermediate reduced light intensity, $\alpha\s{m}$ say.

\subsection{Photo-stationary shapes of sheets}\label{subsect:stationary}
\begin{figure}[!hbt]
\includegraphics[width=\columnwidth]{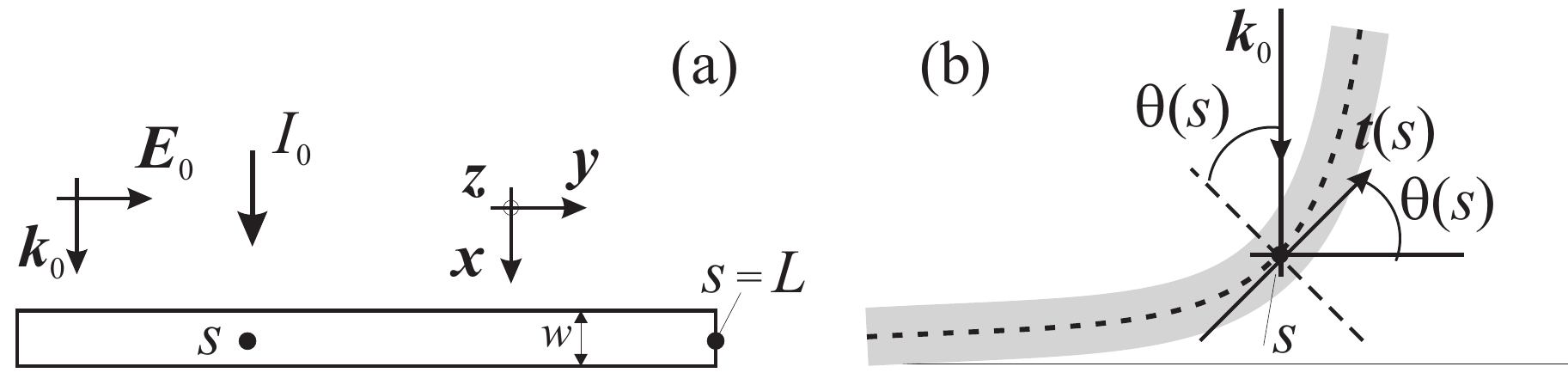}
\caption{A cantilever illuminated from above by light entering the solid with intensity $I_0$. (a) Initially flat, (b) curled up so that the tangent $\vec{t}(s)$ at $s$ makes an angle $\theta(s)$ with $\vec{y}$, and light is incident at angle $\theta(s)$ to the cantilever's normal.}\label{fig:geometry}
\end{figure}

We are concerned with the interplay between bend as a function of incident intensity, but also bend making the sheet oblique to the incident light and thereby itself influencing the effective intensity and penetration of the light responsible for the bend.  This connection and feedback determines the photo-stationary state and also the complex dynamics that we later examine.  To establish and illustrate such qualitative effects, we adopt the simplest possible variation of intensity penetrating the upper surface of the sheet, namely $I_0\cos\theta$ which simply expresses the dilution of the incident flux $I_0$ to an effective flux by obliquity.  This assumption will be deficient in detail since (a) there are complicated Fresnel coefficients governing the wave amplitude refracted into the medium, and (b) the wave entering is not completely refracted to be along the normal to the sheet (though it is nearly so in the case of strong absorption where Snell's law takes an extreme form).  In consequence, the electric vector will not be along the local beam direction, see Fig.~\ref{fig:geometry}(b), and its effect on inducing bend will have another angular factor. Consequently, though our analysis reproduces the qualitative features of self-eclipsing and dynamics dependent on \textit{trans}-depletion front penetration, the precise shapes of sheets to be compared with future experiments may depend on these refinements of angular dependence.

At an arc distance $s$ along the sheet, the tangent and normal are rotated through an angle $\theta(s)$, see Fig.~\ref{fig:geometry}(b).  It is the local $x$ direction (the thickness direction) that enters the attenuation equation (\ref{eq:productlog}) and the effective intensity is $\alpha_0 \cos\theta$, where $\alpha_0= I_0/I_m$ is the effective intensity were the beam to strike normally.  Hence $\theta(s)$ enters the solution ${\mathcal I}$ to be injected into eq.~(\ref{eq:curvature_Iw}) for the local curvature, itself having explicit $\theta$ dependence through the $\alpha_0\cos\theta$ factor appearing.

\begin{figure}[h]
\includegraphics[width=0.8\columnwidth]{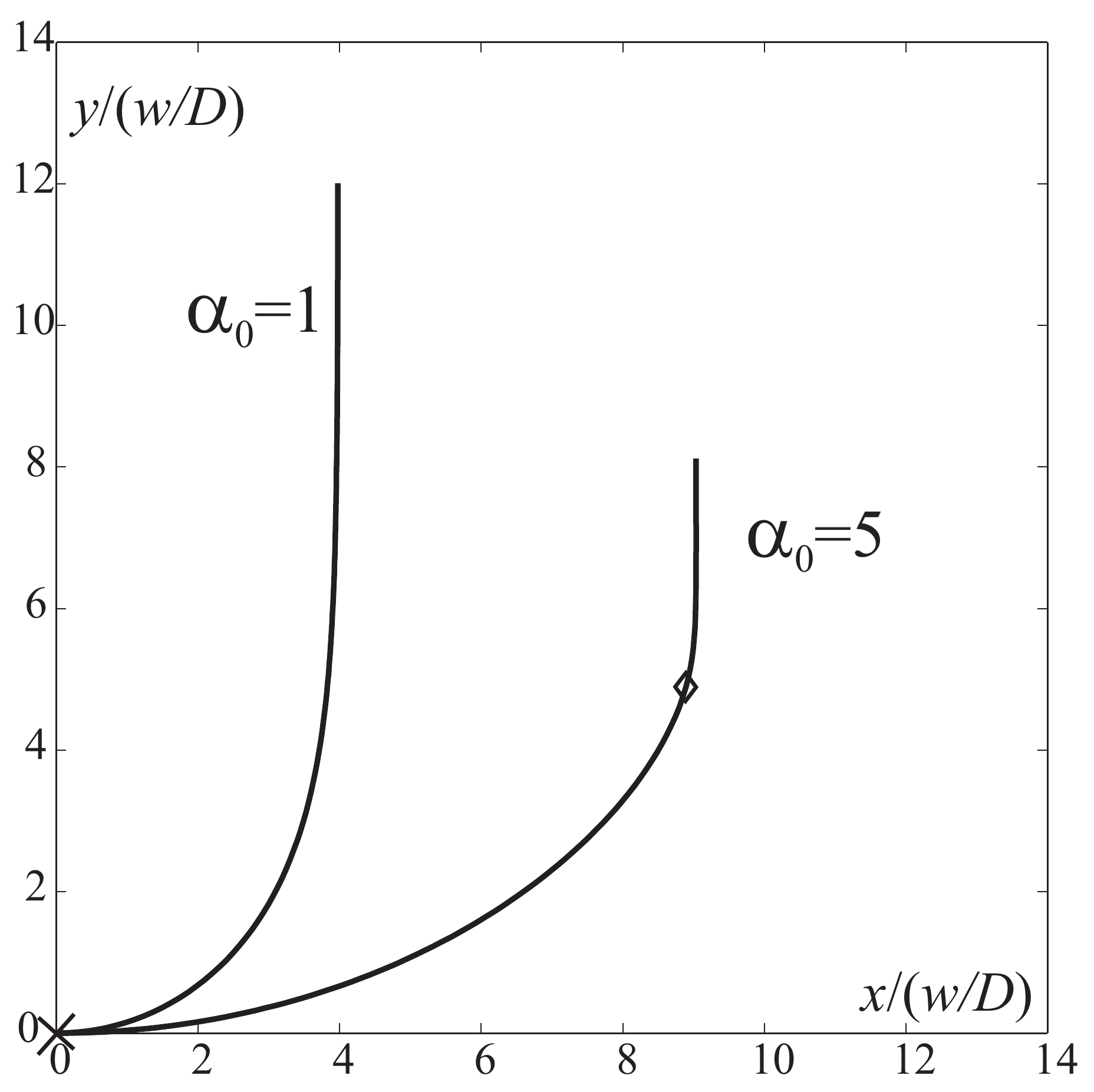}
\caption{Photo-stationary shapes for $\alpha_0=1$ and $\alpha_0=5$. The positions of maximum curvature correspond to angles, and therefore arc positions, marked on Fig.~\ref{fig:curvature-light} ($\times,\diamondsuit$). The reduction of curvature by $w/D$ is used here for lengths too.  Sheets of reduced length $L/(w/D) = 14$ are shown.  For a given $\alpha_0$ and $w/d$, these are master curves; sheets of smaller $L$ will terminate at intermediate places along the curve.}\label{fig:stationary shape}
\end{figure}

Eq.~(\ref{eq:curvature_Iw}) can be written in a superficially simpler form as
\be
\d\theta/\d s=\frac{1}{R}=\frac{D}{w}a \cos\theta(s),\label{eq:curvature cos}
\ee
where there is also $\theta$-dependence in $a$: \be a= \alpha_0[\frac{w}{d}\mathcal{I}_w-(1-\mathcal{I}_w)(1-\frac{w}{2d})-\frac{\alpha_0\cos\theta}{2}(1-\mathcal{I}_w^{2})]\label{eq:a}
\ee both explicitly, and buried in $\mathcal{I}_w$. This differential equation can be integrated to give $\theta(s)$, and then a second integration gives the photo-stationary shape $(x(s),y(s))$ of the sheet bending in the $x-y$ plane. See Fig.~\ref{fig:stationary shape} for shapes corresponding to two effective intensities $\alpha_0$ (that would be falling on flat sheets) that are less than or greater than $\alpha\s{m}$; the two cases are qualitatively different.

Consider incident light with $\alpha_0=1 < \alpha\s{m}$, that is smaller than the $\alpha$ for optimal curvature for this $w/d$. The effective intensity incident locally on the sheet, $\alpha=\alpha_0\cos\theta$, decreases with $\theta$. Therefore for $\alpha_0=1$, the maximum curvature obtains at $s=0$, as shown in Fig.~\ref{fig:stationary shape}, where $\alpha(s)$ is greatest, namely $\alpha_0$. However incident light with $\alpha_0=5$ is more intense than optimal: As $\theta$ increases, $\alpha$ decreases down to $\alpha\s{m}$ (if the total length $L$ is large enough) and curvature increases to a maximum.  If $\theta$ continues to increase, then $\alpha(s)$ and thus curvature decreases to 0. Thus the maximum curvature for $\alpha_0=5$ obtains at an intermediate $s$ in the sheet; see Fig.~\ref{fig:stationary shape}.

To give a simple insight into the effect of curvature being arc-position dependent, one can look for solutions of eq.~(\ref{eq:curvature cos}) for curvature where we take the explicit, leading $\theta$-dependence via $\cos\theta$, and ignore the further $\theta$-dependence in $a$ which we now set to be constant.
Reducing arc lengths by $w/(aD)$, that is $s=u\,w/(aD)$, then $\d\theta/\d u=\cos\theta(u)$ integrates simply to $\sin\theta(u)=\tanh(u)$. Recognising that $(x(u),y(u))=w/(aD)\int_{0}^{u}\d u'(\cos\theta(u'),\sin\theta(u'))$ and using $\cos\theta(u)=\d\theta/\d u$ yields the parametric forms.
\begin{align}
&x(u)=\frac{w}{aD}\theta(u)=\frac{w}{aD}\sin^{-1}(\tanh(u))\notag\\
&y(u)=\frac{w}{aD}\ln(\cosh(u))
\end{align}
and eliminating $u$ gives
\be
y(x)=-\frac{w}{aD}\ln\left(\cos[x/(w/aD)]\right)
\ee
which is qualitatively of the form in Fig.~\ref{fig:stationary shape}.

\subsection{The marginal effect of gravity}\label{sect:gravity}
To rule out any role of gravity in determining static shapes, and later the dynamics of a glass sheet or cantilever, we calculate an extreme bound from Fig.~\ref{fig:geometry}(a). Considering such a sheet of width $W$, we can calculate the curvature close to the clamped end induced by gravity, the torque being estimated as if the sheet were flat ({\it i.e.} estimating the most critical location for curvature and over-estimating the torque). Equating elastic and gravitational torques,
\be
\frac{EWw^3}{12R}=\frac{\rho wWgL^2}{2}
\Rightarrow\frac{w}{R}=\frac{\rho g}{E}\frac{6L^2}{w}\equiv\frac{6L^2}{lw}
\ee
where $l=E/\rho g$ is a characteristic length emerging from matching elasticity ($E$ -- the Young's modulus) with gravity ($g,\rho$ -- acceleration of gravity and density of the photo-glass). For photo-glasses $l\sim10^5$m.

If gravity were to change the beam's curvature by that of a quarter circle of radius $R$, thus $L=\pi R/2$, then $1/R=\pi/2L$ in the above yields a length $L\s{g}=(\pi lw^2/12)^{1/3}$ before gravitational effects compete with elastic.  Ikeda and Yu \cite{Ikeda:03} have $w=7\mu$m and $L\sim3$mm (half their total sample length since they clamped in the middle and not one end), whence $L\s{g}\sim10^{-2}$m is comfortably larger than their $L$. Mol~{\it et al} \cite{Mol:05} have $w\sim10\mu$m -- 40$\mu$m, and their $L=10^{-2}$m is comfortably within the (extremely low) estimates of $L\s{g}=2-4\times10^{-2}$m. We henceforth ignore gravitational effects.

\section{Dynamical photomechanical response and eclipsing}\label{sect:dynamics}
The varying {\it cis}-fraction of dye with depth and time, $n\s{c}(x,t)$, arises from penetration of a {\it trans} depletion front.  Sometimes this front is loosely called a bleaching wave or front, since the converted dye is not an effective absorber of the original colour of light when in the {\it cis} state.  However this bleaching is reversible (not chemical) and is otherwise referred to as saturated absorption \cite{Stryland:00}. The resulting curvature (\ref{eq:curvature}) determined by $n\s{c}$ is thus a function of time, non monotonic since $1/R$ depends on the spatial variation of $n\s{c}(x)$. Thus we have time dependent over-bend, and this is further complicated by (a) the flux driving the depletion front varies with $\theta(s)$, which results from the accumulation of curvature $\partial\theta/\partial s$ from $s'=0$ to $s$, and (b) development of angles $\theta(s,t)>\pi/2$ giving eclipsing, the incident light being blocked from falling on the sheet at arc positions before that $s$ where $\theta=\pi/2$. Such sections are in the dark, their $n\s{c}(x)$ fraction of {\it cis} recovers, and they lose their curvature. The sections of sheet that are doing the eclipsing necessarily have their under sides now exposed to the light and their curvature is reduced and eventually reversed. A reversal can lead to double eclipsing. We now explore this complex dynamics.

The {\it cis} fraction at any time can be taken from (\ref{eq:light absorption}), solving for $n\s{t}$ and replacing it by $1-n\s{c}$.  Thus $n\s{c}=1+(d/\mathcal{I})\,\partial\mathcal I/\partial x=1-d\,\partial A/\partial x$ where the absorption $A=-\ln(\mathcal I)$. When the above $n\s{c}$ is injected into eq.~(\ref{eq:curvature}), the $1$ term in $n\s{c}$ gives a vanishing integral. The gradient term either integrates trivially against $w/2$, or integrates by parts against $-x$ to give overall
\be
\frac{w}{D}\frac{\partial\theta}{\partial s}\equiv\frac{w/D}{R}=\half \frac{w}{d}A(w,t)-\int_{0}^{w}A(x,t)\frac{\d x}{d}.\label{eq:curvature-A}
\ee
We have used $A(0,t)=0$ since $\mathcal I(0,t)=1$.

The coupled, non-linear pair of PDEs (\ref{eq:dynamics}) and (\ref{eq:light absorption}) for $\dot n\s{c}(\equiv-\dot n\s{t})$ and $\mathcal I'$ can be reduced to a single temporal quadrature for $A$ \cite{Corbett:08b} at each $s$:
\be
\dot A=x/d-A+\alpha_0\cos\theta(s,t)(\e{-A}-1),\label{eq:A}
\ee
with $A(0,t)=0$ and $A(x,0)=x/d$. The latter is Beer's law of exponential decay $\mathcal I=\e{-x/d}$ of light in an as-yet undepleted dye population. This 2nd condition needs careful reexamination after eclipsing when sheets start to be irradiated from the back face.

We solve (\ref{eq:A}) (see \cite{Corbett:08b}) for  depletion front solutions and inject them into (\ref{eq:curvature-A}) for $(w/D)/R(s,t)$. The form of the dynamics, especially eclipsing, depends critically on the sheet length $L$ and the reduced intensity $\alpha$. Longer $L$ means that higher angles $\theta(s,t)=\int_{0}^{s}\d s'\partial\theta/\partial s'$ can be accumulated and $\theta\rightarrow\pi/2$ is more achievable. When the tip of the sheet approaches $\pi/2$, it can be convected over to $\theta>\pi/2$ by continuing light penetration at $s<L$ since curvature increases with a further increasing gradient of $n\s{c}(x)$ deeper through the sheet thickness. This non-linearity in response leads to eclipsing before later recovery.
\begin{figure}[h]
\includegraphics[width=0.8\columnwidth]{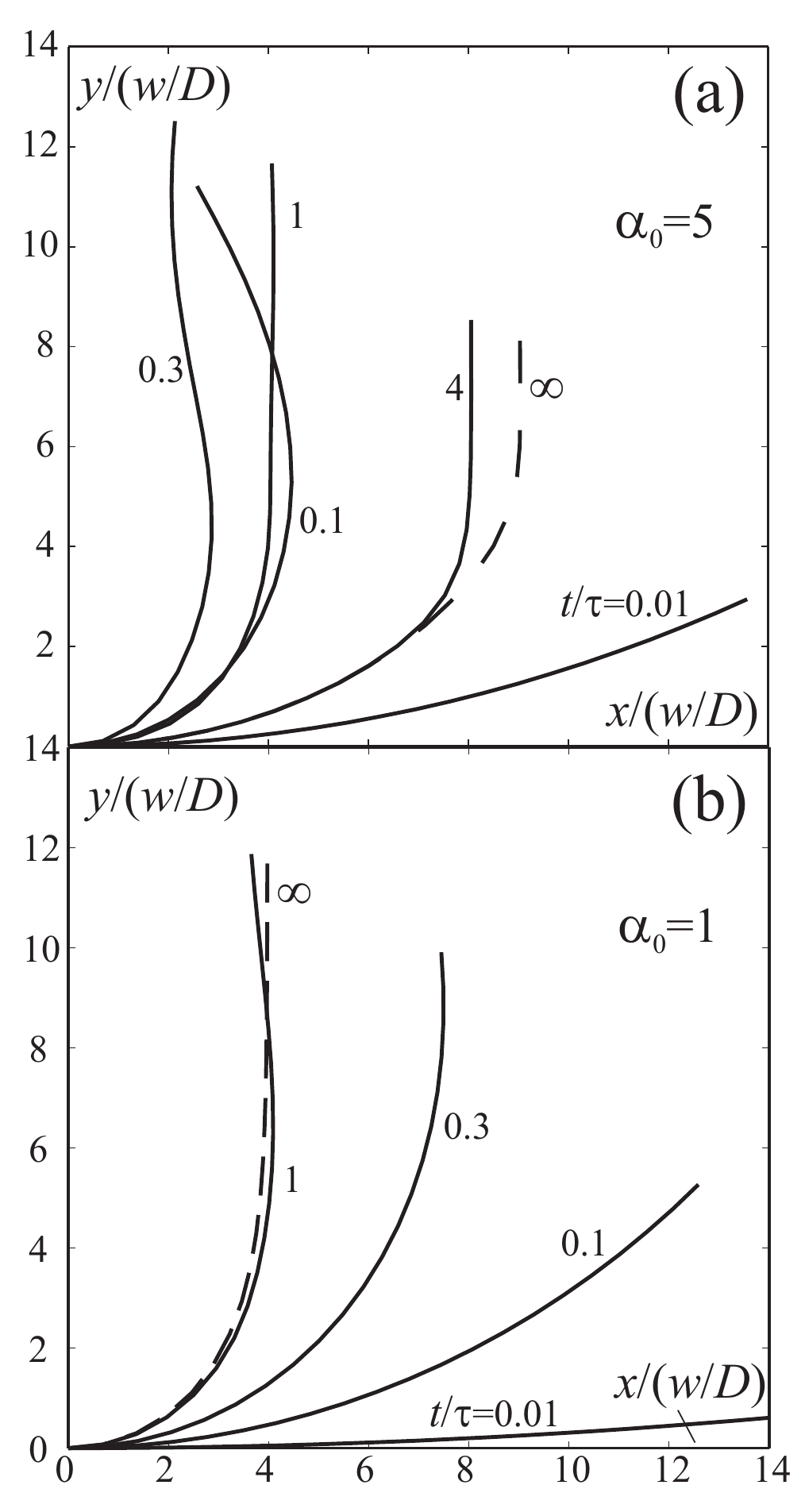}
\caption{Bend curve sequences in $t/\tau$ of 0.01, 0.1, 0.3, 4, $\infty$ for (a) $\alpha_0 = 5$ and (b) $\alpha_0=1$ (where the  $t/\tau = 4$ curve sits on top of that for $t/\tau = \infty$). Reduced sheet length is again $L/(w/D) = 14$, and $w/d = 3$.}\label{fig:dynamics}
\end{figure}

Fig.~\ref{fig:dynamics}(a) shows the dynamics for $\alpha_0=5$. The corresponding photo-stationary state in Fig.~\ref{fig:stationary shape} is the dashed line in Fig.~\ref{fig:dynamics}(a). The sheet starts bending from an initially flat shape. One already sees overshoot at $t/\tau=0.1$. The overshot section is now illuminated on what was the back face, while at least part of the sheet with $\theta(s) < \pi/2$ is eclipsed. The maximum curvature is more or less in the middle of the sheet where it was most strongly bent before being eclipsed. After overshoot, parts with $\theta(s) > \pi/2$ bend backward because of the reversed illumination, as can be seen in $t/\tau=0.3$, while eclipsed parts unbend exponentially in time since they are in the dark. Approach to the stationary state takes a long time, in terms of the fundamental time scale $\tau$, because of the complex sequence of overshoots.

Fig.~\ref{fig:dynamics}(b) has $\alpha_0=1$, which is slightly less than the maximal value $\alpha_m$ for $w/d = 3$.  Now strongest bending is closer to the fixed end, as expected from Fig.~\ref{fig:stationary shape} for this $\alpha_{0}$.  Overshoot is not so extreme and the approach to stationarity is much quicker.  For smaller $\alpha_{0}$, for instance 0.5 for this length $L$, overshoot still occurs, but only slightly, whereas for $\alpha_{0} = 0.1$ with this $L$, it is lost.
\section{Conclusions}\label{sect:conclusions}
We have demonstrated that because the effective illumination is controlled by orientation, and at the same time drives orientation since it induces bend, the bending of photo-responsive sheets is subtle.  To explore the qualitative behaviour that should arise, both in statics and in dynamics, a simple geometrically-inspired dependence of light penetration on angle of incidence is adopted.  The other essential physical driver of this photo-mechanics is that conversion of the isomerising guest dye molecules to their excited state has to be considerable (i) to perturb the local structure of the glass and induce mechanical change, and (ii) to allow deep penetration (via a front of depletion of the ground state species) so that bend can actually occur.  We then find:
\begin{itemize}
  \item Photo-stationary shapes arising are qualitatively different according to whether the intensity of illumination normally incident is more or less than a characteristic value that depends on both a material constant and on the thickness of the sheet reduced by the Beer length for adsorption.  These stationary states cannot be self-eclipsing.
  \item Seminal experiments on nematic glass bend response did display self-eclipsing as response proceeded.  By analysing the dynamical evolution of the mechanics, we show that this must have arisen as a transient effect and, accordingly, must be impossible to explain with just linear, Lambert-Beer, light absorption.  We further predict that the route to the final photo-stationary state must show partial unbending via back bend, and could display multiple eclipses.
\end{itemize}

The two regimes both suggest further experiments which should also
give insight into how complex the true angular dependence of
absorption is.  For example the experiments of Ikeda and Yu
\cite{Ikeda:03} could be illuminated for longer, such that appreciable
backbend (in addition to eclipsing) is observed.  Eclipsing itself
depends quite crucially on the length of the photo-responsive sheets,
therefore repeating the experiments of Ikeda and Yu with different
sample lengths should also reveal a complicated behaviour.   Perhaps most simple of all would be to coat the back face of the sheet with a reflective coating, thus removing effects from illuminating the back surface and allowing one to focus entirely on front-surface illumination and eclipsing.

\vspace{.3cm}
\noindent \textit{Acknowledgements.}  DC is grateful for support from BBSRC and XC from the Chinese Government visiting student programme. We thank Professor Yu for the photograph of a self-eclipsing photo-responsive nematic glass sheet.  DC would like to dedicate this paper to the memory of Tess Tracey.%merlin.mbs apsrev4-1.bst 2010-07-25 4.21a (PWD, AO, DPC) hacked
%Control: key (0)
%Control: author (8) initials jnrlst
%Control: editor formatted (1) identically to author
%Control: production of article title (-1) disabled
%Control: page (0) single
%Control: year (1) truncated
%Control: production of eprint (0) enabled
%.
%\bibliography{references-photo-bending}

% Produces the bibliography via BibTeX.
%paste in by hand (from photo_eclipsing.bbl) the prepared bibliography when ready and comment out the line

\end{document}